\documentclass[letterpaper, 10pt, conference]{ieeeconf}  % Comment this line out if you need a4paper
\usepackage{graphicx}
\usepackage{amsmath,amssymb,color,bm,cite}

\IEEEoverridecommandlockouts                              % This command is only needed if
 \newtheorem{theorem}{Theorem}[section]
 
 \newtheorem{lemma}[theorem]{Lemma}

 \newtheorem{remark}[theorem]{Remark}
  \newtheorem{assumption}[theorem]{Assumption}
    \newtheorem{observation}[theorem]{Observation}
\begin{document}

\title{\LARGE \bf The Social System Identification Problem}

\author{Hoi-To Wai$^{\dagger}$, Anna Scaglione$^{\dagger}$, Amir Leshem$^{\ddagger}$
\thanks{*This material is based upon work supported by NSF CCF-1011811.}% <-this % stops a space
\thanks{$^{\dagger}$H.-T.~Wai and A.~Scaglione are with School of Electrical, Computer and Energy Engineering, Arizona State University, Tempe, AZ 85281, USA.
               Emails: {\tt \footnotesize \{htwai,Anna.Scaglione\}@asu.edu}}
\thanks{$^{\dagger}$A.~Leshem is with Faculty of Engineering, Bar-Ilan University, Ramat-Gan, Israel.
              Email:  {\tt \footnotesize leshema@eng.biu.ac.il}}
}

%%% ----------------------------------------------------------------------

%%% ----------------------------------------------------------------------
\maketitle
%%% ----------------------------------------------------------------------
\begin{abstract}
%There is a vast literature focused on modeling opinion diffusion in social networks capturing the dynamics of agents opinions due to their interactions. Fewer have tried to fit these models on observed network dynamics.
The focus of this paper is modeling what we call a \emph{Social Radar}, i.e.  a method to estimate the  relative influence between social agents, by sampling their opinions and as they evolve, after injecting in the network  stubborn agents. The stubborn agents opinion is not influenced by the peers they seek to sway, and their opinion bias is the known input to the social network system. The novelty is in the model presented to probe a social network and the solution of the associated regression problem. The model allows to map the observed opinion onto system equations that can be used to infer the social graph and the amount of trust that characterizes the links.
\end{abstract}

\section{Introduction}\label{intro}
Recently, the rapid growth of online social media such as Facebook, Twitter has generated a lot of interests in researches on social networks. Importantly, it has provided a platform for researchers from multiple disciplines, ranging from social science to statistical physics, to study and understand the behavior of the human society.
In the controls area, due to the close ties with decentralized robots coordination problems,  several works focuses on modeling the opinion dynamics/exchanges (see e.g. \cite{acemoglu,Acemoglu2010,Jia2013,Ravazzi2015,TahbazSalehi2008,Touri2011,Carli2010,Blondel2005} and references therein). Arguably, understanding the opinion dynamics through which individuals seek to share information and agree is one of the most important social studies.
In light of this, extensions to the basic opinion dynamics model are also popular, e.g.,
\cite{Deffuant,Hegselmann2002,yildiz2010computing,yildiz2011discrete,Blondel2009,li2013consensus}.
Also, in the recent developments, the notion of controllability and observability has been extended to the context of complex networks \cite{Wang2014,Doostmohammadian2014,Liu2011,Liu2013}.

The common definition of a social network focuses on the \emph{social graph} component \cite{Jackson2008}, where we represent individuals  (social agents) as \emph{nodes} and the friendships between them as \emph{edges}.
Knowing the social graph alone is insufficient for understanding social networks.
In particular, individuals may exhibit different degree of \emph{trust} in their neighbors. There is strong trust among the close friends and weaker trust between individuals without mutual interests \cite{degroot,Friedkin1990}. Identifying the social system  allows us to predict the behavior of individuals in a  social network in times of decision making.
 The difficulty is that, while the interaction between agents may be evident, {\it the trust between them and the impact an interaction has on another agent is not directly observable}.
% and will affect the outcomes of opinion dynamics processes.
%, where individuals in the social network seek consensus by communicating to the others.

%Our study paves the way towards designing a \emph{social radar}.

The focus of this study is identifying what we call a \emph{social system}, which encompasses  both the \emph{social graph} and the set of \emph{trusts between individuals}.
We define
%the \emph{input} as the initial opinions held by the individuals;
the \emph{state} as the  instantaneous opinions; and the \emph{output} as the observed opinions after a certain period of time of discussions. 
The social system is modeled as an \emph{endogenous} system of opinion updates that follows exogenous stimuli that we observe.

We are interested in solving the \emph{social system identification (SSI)} problem, which could be viewed as the development of a \emph{Social Radar}
since the idea is injecting \emph{test signals} into the system and observing its output \cite{Soderstrom1988}, much like in a traditional Radar. An inverse problem is then solved by gathering observations that are tied to the input opinions and output opinion pairs. The challenge is that the latter are endogenous.
%The acquired information will be useful, for example, for enhancing a certain promotion campaign.
%In particular, we are interested in The focus of this paper is to propose a method for identifying such a social system.
In the literature, a related issue is the inference problem of graphical models, e.g., \cite{Wainwright2008,Anandkumar2012,Bresler2014}.  The latter assumes that the effect of the social system is manifested in the correlation of the neighboring opinions, instead of the opinion dynamics.
%A related issue of observability has been covered recently in \cite{Liu2011,Liu2013}, which focused on finding the minimal set of sensors in order to observe the entire state by assuming prior knowledge on the social graph.
Indeed, our model is similar to that considered in  \cite{De2014,Timme2007,Wang2011b}. Particularly, the methods proposed in \cite{Timme2007,Wang2011b} consider the case with non-linear dynamics. However, the methods require knowing precisely when the opinion of an agent has impacted that of another, which we claim is unrealistic, given that what happens in people brains is not visible and tracking their impact
on an individual would require testing the opinion of the neighborhood again, which is an unnatural way of communicating.

In contrast, our work assumes only partial knowledge on the social graph and that opinion updates have occurred at unknown times due to unknown stimuli. All that we observe are noisy versions of the agents opinions. This motivates us to use steady state models
as an approximation for what ties the opinions we sample over time.
However, to identify the system we need, therefore, to prevent trivial consensus. Our idea is to introduce a set of stubborn agents, i.e., agents who are not swayed by other opinions \cite{yildiz2010computing,yildiz2011discrete,Acemoglu2013,bianchi2012},  into the social network. The stubborn agents serve as `probes' inserted on the social network that injects \emph{input} to a social system. Indeed, the stubborn agents  change the terminal behavior of the opinion dynamics and reveal the social system in the form of an underdetermined linear system.

The contributions of this paper are two-fold. Firstly, we formulate the SSI problem under the presence of stubborn agents, and provide a set of conditions for identifiability. Secondly, we consider the random opinion dynamics model and propose an estimator for the (ensemble) mean of terminal opinions. The mean square  convergence of such estimator is proven.
We provide numerical results to verify our findings.

\textbf{Notations}: The Kronecker product is $\otimes$ and ${\rm vec}(\cdot)$ as the vectorization operator. Moreover,  ${\rm Diag}: \mathbb{R}^n \rightarrow \mathbb{R}^{n \times n}$ and ${\rm diag}: \mathbb{R}^{n \times n} \rightarrow \mathbb{R}^{n}$ are defined as the diagonal operators on square matrices and vectors, respectively.

% then we talk about stubborn agents...

%An important question that has left unanswered is the \emph{identifiability} of a social system.
%
%This work focuses on the design of a \emph{social radar}.
%
%
%The study of social networks stems its root from social science

\section{System Model}

The social network has an associated graph $G = (V,E)$ where the vertexes are the set of agents  $V = [n] \triangleq \{1,2,...,n\}$ and 
\emph{social system} we want to explore is  the tuple $S = (E, \overline{\bm W})$ where $\overline{\bm W}$ is the
 row-stochastic matrix of the trust coefficients. 
We assume: \vspace{.1cm}
\begin{assumption} \label{assume:E}
The trust matrix $\overline{\bm W}$ satisfies $\overline{W}_{ij} > 0$ and/or $\overline{W}_{ji} > 0$ if and only if $ij \in E$.
$\overline{W}_{ii}$ is the self-trust. \vspace{.1cm}
\end{assumption}
Our goal is to identify $\overline{\bm W}$ by observing opinions whose dynamics are consistent with $S$.
Specifically, we assume that the agents are shaping opinions over a certain issue $s \in \mathbb{N}$. Initially, each agent holds an opinion (belief) on a discrete random variable $\Theta_s \in \{\theta_1,\ldots,\theta_m\}$,  i.e. the p.m.f.~${\bf x}_i(0;s) = ( p(\theta_1|s_{i,s}), p(\theta_2|s_{i,s}),..., p(\theta_{m-1} |s_{i,s}) )^T$, where $p(\theta_j |s_{i,s})$ is the $i$th agent's belief on the event $\Theta_s = \theta_j$ and $s_{i,s}$ is the \emph{private information} agent $i$ has before its interactions. Notice that   $p(\theta_{m} |s_{i,s}) = 1 - {\bf 1}^T {\bf x}_i(0;s)$.
The beliefs are forged by the DeGroot's model \cite{degroot}:
\begin{equation} \label{eq:opn_dyn}
{\bf x}_i(t+1;s) = \sum_{j \in {\cal N}_i} W_{ij}(t) {\bf x}_j(t;s),
\end{equation}
where ${\cal N}_i$ is the set of neighbors of $i$ and $W_{ij}(t)$ is the $(i,j)$th element of the row stochastic matrix ${\bm W}(t)$. We assume that ${\bm W}(t)$ is i.i.d.~and drawn from a p.d.f.~satisfying $\mathbb{E} \{ {\bm W}(t) \} = \overline{\bm W}$.
The opinion dynamics can be described as
\begin{equation} \label{eq:opn}
{\bm x}(t+1;s) = {\bm W}(t) {\bm x}(t;s),
\end{equation}
where $\bm{x}(t;s) = ({\bf x}_1(t;s),...,{\bf x}_n(t;s))^T$ stacks the $1 \times m$ vectors to form an $n \times m$ matrix. Notice that the above dynamics includes the \emph{randomized models} in \cite{Boyd2006,Aysal2009} as special cases.
Moreover, what is \emph{active} at a given time is random and the trust matrix $\overline{\bm W}$ is embedded in the opinion dynamics \eqref{eq:opn}. 
Our observations are actions/ratings that an agent performs and that the social network is exposed to (e.g., `liking' a post on Facebook). We assume it is possible to take a noisy snapshot of the opinions at time $t_i$:
\begin{equation} \label{eq:noisy_meas}
{\bm y}(t_i;s) = {\bm x}(t_i;s) + {\bm n}(t_i;s),
\end{equation}
where ${\bm n}(t_i;s)$ contains i.i.d.~noise samples with bounded variance $\sigma^2$. It is due to the fact that we do not have direct access to the opinion or are using incomplete or outdated information about it. Eq.~\eqref{eq:opn} \& \eqref{eq:noisy_meas} give a linear system representation for the social system $S$ with the state being the opinions. 
%We assume full observability on the system state.

We define the \emph{social system identification (SSI)} problem as the task of inferring $\overline{\bm W}$ from a set of  measurements ${\bm y}({\cal T}_s; s) \triangleq \{ {\bm y}(t_i;s) \}_{ t_i \in {\cal T}_s }$ (and over a number of issues $s=1,...,K$).
The general SSI problem is challenging to solve for several reasons. For example, we see that $\overline{\bm W}$ is hidden in the random model \eqref{eq:opn} and \eqref{eq:noisy_meas}; also, from \eqref{eq:opn} it is not even possible to retrieve ${\bm W}(t)$ from the samples since $m \ll n$, i.e., the linear equation is rank-deficient. We see that additional prior knowledge must be incorporated to develop a tractable SSI method.

There are a few prior studies on the SSI problem. Most closely related to ours is the work in \cite{De2014,Timme2007,Wang2011b}.
In particular, \cite{De2014} considers the same model as ours. The authors assume that the set ${\cal T}_s$ is consecutive, e.g., ${\cal T}_s = \{ t_i, t_i+1,..., t_j \}$ and the trust matrix is static ${\bm W}(t) = \overline{\bm W}$ with \emph{known} sparsity pattern. In \cite{Timme2007,Wang2011b}, the authors consider a nonlinear dynamical system and applied compressed sensing to infer the network topology from  samples marked with time stamps.

%The assumptions made in \cite{De2014} maybe unrealistic in several applications, for example in the development of a \emph{social radar}. In particular, the topology of the social graph $E$ may be unknown or (at best) partially known to a social radar and the trust matrix ${\bm W}(t)$ should not be static since social interactions are asynchronous and random.

\begin{figure}[t]
\centering
 \includegraphics[width =0.8\linewidth]{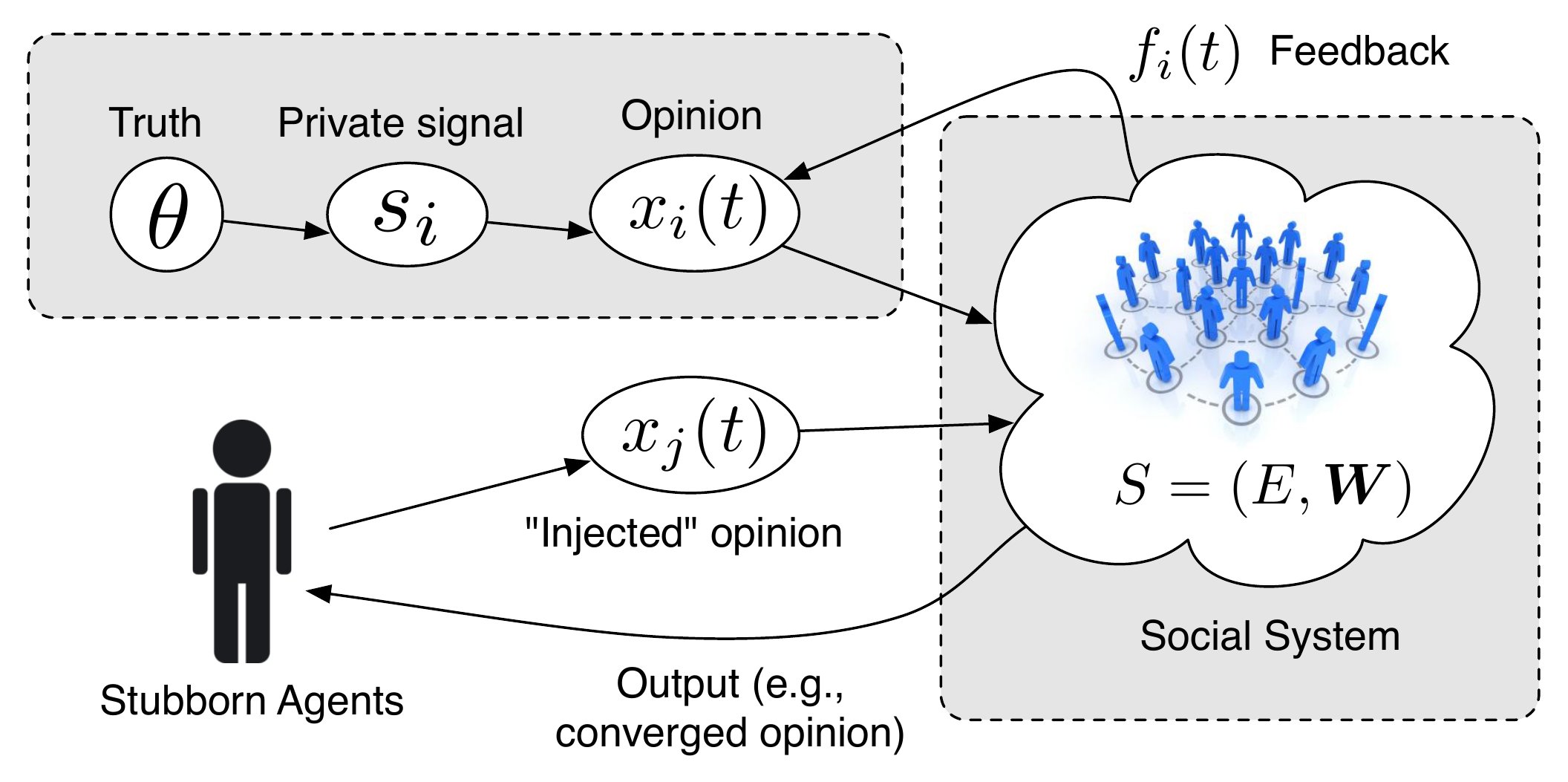}\vspace{-.2cm}
\caption{\small Using stubborn agents as `probes' on social systems} \vspace{-.6cm}
\label{fig:graph}
\end{figure}

The assumptions made in \cite{De2014,Timme2007,Wang2011b} may be restrictive for social networks as the rate of interaction is unknown for the latter, therefore the time stamp information cannot be obtained accurately.
Our idea is to introduce a set of \emph{stubborn agents}  as \emph{probes}; see Fig.~\ref{fig:graph}. The developed SSI method requires only partial knowledge on the topology $E$ and is applicable to the scenario with time-varying trust matrix.

%remainder of this paper will be devoted to the development of such SSI methods.

\section{Stubborn Agents} \label{sec:stub}

In social networks, \emph{stubborn agents} are those members who place \emph{zero trust} on their neighbors. In our framework we 
assume there are $n_s < n$ {stubborn agents} in the social network that we know of or control. We index them by $V_s = [n_s]$.
The resulting trust matrix ${\bm W}(t)$ is:
\begin{equation} \label{eq:stubborn_w1}
{\bm W}(t) = \left( \begin{array}{cc}
{\bm I}_{n_s} & {\bm 0} \\
{\bm B}(t) & {\bm D}(t)
\end{array}\right),
\end{equation}
and similar structure can be found in $\overline{\bm W}$. 
%Notice that ${\bm W}(t)$ is  row-stochastic for all $t$.
%If we assume that the connectivity of the stubborn agents $E(V_s) = \{ xy : x \in V_s, y \in V \setminus V_s \}$ is known then we know the sparsity pattern of $\overline{\bm B}$ (and ${\bm B}(t)$), i.e., ${\cal E}_{B} = \{ ij : \overline{B}_{ij} > 0 \}$. 
We assume:\vspace{.1cm}
%\begin{assumption}
%The lower right block, $\overline{\bm D}$, is full rank and satisfies $\| \overline{\bm D} \|_2 < 1$.
%\end{assumption}
\begin{assumption} \label{assume:b}
The support of $\overline{\bm B}$, ${\cal E}_{B}  = \{ ij : \overline{B}_{ij} > 0 \} = E(V,V_s)$, is known. Moreover, each agent in $V$ has non-zero trust on at least one agent in $V_s$.\vspace{.1cm}
\end{assumption}
The assumption asserts that each non-stubborn agents is influenced by at least one stubborn agents. 
As the graph $G$ is connected, Assumption~\ref{assume:b} implies that the principal submatrix ${\bm D}$ satisfies $\| {\bm D} \|_2 < 1$. 

%\begin{assumption} \label{assume:b}
%The lower left block, $\overline{\bm B}$ is full rank, i.e., ${\rm rank}(\overline{\bm B}) = \min\{ n_s, n-n_s\}$. \vspace{.1cm}
%\end{assumption}
%Assumption~\ref{assume:b} is similar to that each of the stubborn agents is influencing at least an independent non-stubborn agent. 
%Notice that it also implies $\| \overline{\bm D} \|_2 < 1$. 

We demonstrate the specific structure in \eqref{eq:stubborn_w1} gives rise to a set of equations that allows a tractable solution to SSI.
% talk about the
To begin with, we assume that the opinion exchange is static, i.e., ${\bm W}(t) = \overline{\bm W}$ for all $t$; this assumption will be relaxed later in Section~\ref{sec:rand}. Observe the following:
\begin{observation} \cite{yildiz2010computing}
Under the assumption that ${\bm W}(t) = \overline{\bm W}$. Consider   \eqref{eq:opn} by setting $t \rightarrow \infty$, we get:
\begin{equation}
\lim_{t \rightarrow \infty} {\bm x}(t;s) = \overline{\bm W}^\infty {\bm x}(0;s),
\end{equation}
where
\begin{equation} \label{eq:stubborn_w}
\overline{\bm W}^\infty = \left( \begin{array}{cc}
{\bm I}_{n_s} & {\bm 0} \\
( {\bm I} - \overline{\bm D})^{-1} \overline{\bm B} & {\bm 0}
\end{array}\right). \vspace{.2cm}
\end{equation}
\end{observation}
%Note that ${\rm rank}(\overline{\bm W}) = \min\{n_s,n-n_s\}$. 
%In contrast, when $n_s \leq 1$, the limiting product \eqref{eq:stubborn_w} becomes a rank-one matrix.

Consequently, non-stubborn agents' opinions satisfy:
\begin{equation} \label{eq:limit_eq}
\begin{split}
\lim_{t \rightarrow \infty} ( {\bm I} - \overline{\bm D}) {\bm x}_{[n] \setminus [n_s]} (t;s) & = \lim_{t \rightarrow \infty} \overline{\bm B} {\bm x}_{[n_s]} (t;s) \\
& = \overline{\bm B} {\bm x}_{[n_s]} (0;s),
\end{split}
\end{equation}
where we have defined ${\bm x}_{ [n_s] }(t;s) \triangleq ( {\bf x}_i (t;s) )_{i \in [n_s]}$. In fact, the right hand side of \eqref{eq:limit_eq} can be replaced by $ \overline{\bm B} {\bm x}_{[n_s]} (\tau;s)$ for any $\tau \in \mathbb{N}$ as the stubborn agents never change opinions.

The final opinions are driven by the initial opinions at the stubborn agents. Importantly, Eq.~\eqref{eq:limit_eq} gives a set of linear equations that characterizes the social system $S$. To this end, we can define an estimator for $\lim_{t \rightarrow \infty} {\bm x} (t;s) $:
\begin{equation} \label{eq:sample}
\hat{\bm x} ({\cal T}_s;s) = \frac{1}{ | {\cal T}_s | } \sum_{ t_i \in {\cal T}_s } {\bm y} (t_i; s),
\end{equation}
where the sampling set ${\cal T}_s$ is defined as
\begin{equation}
{\cal T}_s = \{ t_i : i \in [| {\cal T}_s |],~t_i \geq T_o,~\forall~i \},
\end{equation}
where $T_o \gg 0$. Notice that the time indices $t_i$ nor their orders are not required in the computation of \eqref{eq:sample}, i.e., we do not need to know the exact time in which opinion updates have occurred when sampling ${\bm y}(t_i;s)$.
%such that $t_i \neq t_j$ for $i \neq j$, $t_i \geq T_o$, $T_o \gg 0$ for all $t_i \in {\cal T}_s$.

In the case of static exchange, it is easy to check that the estimator \eqref{eq:sample} is consistent as $|{\cal T}_s| \rightarrow \infty$ and $T_o \rightarrow \infty$; see Section~\ref{sec:rand} for a further discussion of its convergence properties. Notice that unlike \cite{De2014}, we do not require the sampling set ${\cal T}_s$ to be composed of consecutive indices.
Consider collecting the estimate \eqref{eq:sample} for $K$ issues (i.e., $s=1,...,K$) into data matrices, we have the  linear equation:
\begin{equation} \label{eq:stub_sys}
( {\bm I} - \overline{\bm D}) {\bm Y} = \overline{\bm B} {\bm Z} + {\bm N},
\end{equation}
where
\begin{equation} \begin{split}
{\bm Y} & \triangleq ( \hat{\bm x}_{[n] \setminus [n_s]} ({\cal T}_s;s) )_{s=1}^K \in \mathbb{R}^{(n-n_s) \times Km} \\
{\bm Z} & \triangleq ( \hat{\bm x}_{[n_s]} ({\cal T}_s;s) )_{s=1}^K \in \mathbb{R}^{n_s \times Km}
\end{split} \end{equation}
denote the data matrices for the opinions at the normal agents and stubborn agents, respectively, and ${\bm N}$ is the additive noise with variance $\sigma_n^2 $ that captures the estimation error from \eqref{eq:sample}.

\subsection{Identifying the Social System} \label{sec:ssi}
Our next endeavor is to formulate the respective inverse problem for SSI. In particular, our goal is to find the tuple $(\overline{\bm B}, \overline{\bm D})$ that satisfies the system of equations:
\begin{equation} \label{eq:sys_eq}
( {\bm I} - \overline{\bm D}) {\bm Y} = \overline{\bm B} {\bm Z} + {\bm N},~(\overline{\bm B}~\overline{\bm D}) {\bf 1} = {\bf 1},~\overline{\bm B}, \overline{\bm D} \geq {\bm 0}.
\end{equation}
We observe the following:
\begin{lemma} \label{lem:amb}
Assume ${\bm N} = {\bm 0}$, there exists a tuple $({\bm B}, {\bm D})$ that satisfies \eqref{eq:sys_eq}. The tuple $ ({\bm B}', {\bm D}')$ also satisfies \eqref{eq:sys_eq} with \vspace{-.3cm}
\begin{subequations} \label{eq:equiv}
\begin{align}
\bm{B}' = \bm{\Lambda} {\bm B},~
{\rm off}(\bm{D}') = \bm{\Lambda} {\rm off}({\bm D}),\\
{\rm diag}(\bm{D}') = {\bf 1} - \bm{\Lambda} ( {\bm B} {\bf 1} + {\rm off}({\bm D}) {\bf 1} ),
\end{align}
\end{subequations}
where ${\rm off}({\bm D})$ denotes the square matrix with only off-diagonal elements in ${\bm D}$
and $\bm{\Lambda}$  is any non-negative diagonal matrix such that ${\rm diag}(\bm{D}') \geq {\bm 0}$.
% for all $\alpha \geq 0$ such that ${\rm diag}(\bm{D}') \geq {\bm 0}$.
\end{lemma}

\emph{Proof}: The existence of $({\bm B}, {\bm D})$ is ensured by picking ${\bm{B}} = \overline{\bm B}, {\bm{D}} = \overline{\bm D}$.
It is also obvious that the second equation in \eqref{eq:sys_eq} is satisfied by $(\bm{B}',\bm{D}')$ for an arbitrary diagonal matrix $\bm{\Lambda}$.
For the first equation  in \eqref{eq:sys_eq}, we observe that
\begin{equation}
\begin{array}{l}
{\bm D}' {\bm Y} + {\bm B}' {\bm Z} = ( {\rm Diag}( {\rm diag}({\bm D}') ) + {\rm off}( {\bm D}') ) {\bm Y} + {\bm B}' {\bm Z} \\
= {\bm Y} - \bm{\Lambda} ( ({\rm Diag}( {\bm B}{\bf 1} + {\rm off}({\bm D}) {\bf 1})  - {\rm off}({\bm D})) {\bm Y} - {\bm B}{\bm Z}) \\
= {\bm Y} - \bm{\Lambda} ( ({\rm Diag}( {\bf 1} - {\rm diag}({\bm D}))  - {\rm off}({\bm D})) {\bm Y} - {\bm B}{\bm Z}) \\
= {\bm Y} - \bm{\Lambda}  ( {\bm Y} - {\bm D} {\bm Y} - {\bm B}{\bm Z}) = {\bm Y},  \vspace{-.2cm}
\end{array}
\end{equation}
where the third equality is due to ${\bm B}{\bf 1} + {\rm off}({\bm D}) {\bf 1} = {\bf 1} - {\rm diag}({\bm D})$.  \hfill \textbf{Q.E.D.}

We define an equivalent relation $\sim$ as:
\begin{equation}
({\bm B}, {\bm D}) \sim ({\bm B}', {\bm D}') : \exists \bm{\Lambda} \geq {\bm 0}~{\rm s.t.}~\eqref{eq:equiv}~{\rm holds}.
\end{equation}
The \emph{relative trust} weights, defined as $D_{ij}^r = D_{ij} / (1 - D_{ii})$ and $B_{ij}^r = B_{ij} / (1 - D_{ii})$,  is preserved for all tuples belonging to the same equivalence class. Moreover, there exists $(\bm{B}',{\bm D}')$ with ${\rm diag}({\bm D}') = {\bm 0}$  such that $(\overline{\bm B}, \overline{\bm D}) \sim (\bm{B}',{\bm D}')$\footnote{The corresponding $\bm{\Lambda}$ can be found as $\bm{\Lambda} = {\rm Diag}(1/\overline{d}_1,\ldots,1/\overline{d}_{n-n_s})$ with $\overline{\bm d} = \overline{\bm B} {\bf 1} + {\rm off}(\overline{\bm D}) {\bf 1}$.}.
Hence, we adopt a pragmatic approach to remedy the scaling ambiguity by fixing ${\rm diag}(\hat{\bm{D}}) = {\bm 0}$.

%In particular, Eq.~\eqref{eq:sys_eq} defines a manifold ${\cal M}_S$ for the social system that captures the \emph{relative trusts} between an agent and his/her neighbors. To this end, we ensure the uniqueness of solution by imposing
%${\rm diag}(\hat{\bm{D}}) = {\bm 0}$.

%The SSI problem thus
%However, in SSI, we are interested in learning the \emph{relative} trusts among agents. To this end, it suffices to add the constraint that
%As a remedy, we limit

The next ingredient is the fact that a social graph has typically a sparse set of links between agents, i.e., $\overline{\bm D}$ is sparse. This motivates us to consider the following $\ell_0$ minimization \cite{Candes2005}  problem:
\begin{subequations} \label{eq:ssi}
\begin{align}
%\begin{array}{rl}
\displaystyle \min_{ \hat{\bm B}, \hat{\bm D} } &~~ \| {\rm off} (\hat{\bm D}) \|_0 \\
{\rm s.t.} & ~~\| ({\bm I} - \hat{\bm D}) \bm{Y} - \hat{\bm B} {\bm Z} \|_F^2 \leq \epsilon, \label{eq:first_cons} \\
& ~~\big( \hat{\bm B}~\hat{\bm D} \big) {\bf 1} = {\bf 1},~\hat{\bm D} \geq {\bm 0},~\hat{\bm B} \geq {\bm 0},\\
& ~~ \hat{B}_{ij} > 0,~\forall~ij \in {\cal E}_B,~\hat{D}_{ii} = 0,~\forall~i, \label{eq:last_cons}
%& ~~\hat{D}_{ij} = \hat{D}_{ji},~\forall~i,j, \label{eq:last_cons}
%\end{array}
\end{align}
\end{subequations}
where $\epsilon > 0$ is a regularization parameter that depends on $\sigma_n$. Notice that the last constraint is due to Assumption~\ref{assume:E} and the prior knowledge on ${\cal E}_B$.

Problem \eqref{eq:ssi} is non-convex. However, Problem \eqref{eq:ssi} can be readily convexified by replacing the $\ell_0$ norm in the objective function by an $\ell_1$ norm:
\begin{equation} \label{eq:ssi_r}
\min_{ \hat{\bm B}, \hat{\bm D} } \| {\rm off}( \hat{\bm D} ) \|_1~{\rm s.t.}~\text{Eq.~\eqref{eq:first_cons} to \eqref{eq:last_cons}}.
\end{equation}
% and (ii) ignoring the last constraint \eqref{eq:last_cons}.
The convexified problem can be solved using off-the-shelf softwares, e.g., CVX \cite{cvx}. 
% Efficient solution methods to the problem will be discussed in our future works.

Next, we study conditions under which \eqref{eq:ssi} can identify the social system $S$.
Intuitively, we see that the identifiability condition depends on the number of stubborn agents and the sparsity of the trust matrix ${\bm D}$.  
% Before we move on, a quick observation on \eqref{eq:first_cons} to \eqref{eq:last_cons} may suggest that by increasing $K$ indefinitely, one can always recover $(\overline{\bm D}, \overline{\bm B})$ as it generates more observations. 
%As the following discussion reveals, there is a limit on $n_s$ to achieving perfect recovery.
Importantly, in the case with \emph{optimized placement of stubborn agents}\footnote{Such is possible in a controlled experiment setting, where the stubborn agents can be controlled to influenced a sub-group of ordinary agents.}, we have the following condition, whose proof can be found in the extended version of this paper \cite{tsipn_submit}. 
\begin{theorem} \label{thm:cs}
Let $n > n_s$ and the support of $\overline{\bm B} \in \mathbb{R}^{n \times n_s}$ be constructed such that each row  has $d$ non-zero elements, selected randomly and independently. 
Define $b_{min} = \min_{ij \in {\rm supp}({\bm B}^r)} {B}_{ij}^r$, $b_{max} = \max_{ij \in {\rm supp}({\bm B}^r)} {B}_{ij}^r$, $n_s = \beta n$, $\beta' = \beta - d/n$ and $\delta = 1 - 1 / (d-1)$. If 
\begin{equation} \label{eq:thmcs}
d > \max \Big\{ 4, 1 + \frac{H(\alpha) + \beta' H(\alpha/\beta')}{\alpha \log (\beta' / \alpha)} \Big\},
\end{equation}
%and 
\begin{equation} \label{eq:thmval}
b_{min} (2 d - 3) - 1 - 2 b_{max} > 0,
\end{equation}
where $H(x)$ is the binary entropy function, 
and $ \| {\bm d}_i^r \|_0 \leq \alpha n / 2$ for all $i$, where ${\bm w}_i^r$ is the $i$th row of ${\bm D}^r$,  then as $n \rightarrow \infty$, solving  \eqref{eq:ssi} yields $({\bm B}^\star, {\bm D}^\star) = ({\bm B}^r, {\bm D}^r)$. \vspace{.1cm}
%${\cal F}$ is a singleton with high probability. In other words, ${\cal F} = \{ (\overline{\bm B}', \overline{\bm W}' ) \}$ is the singleton.
%The failure probability (when $({\bm B}^\star, {\bm D}^\star) \neq ({\bm B}^r, {\bm D}^r)$) is:
%\begin{equation} \label{eq:prob}
%{\rm Pr}( Fail ) \leq \left( \frac{d}{\beta} \right)^4 \frac{d-1}{n^2} + {\cal O}( n^{2- (d-1)(d-3)} ).\vspace{.2cm}
%\end{equation}
\end{theorem}

Using \eqref{eq:thmcs} it is possible to derive a lower bound $\beta(d,\alpha)$ on $\beta$ that depends on $d,\alpha$  and thus the number of stubborn agents required. 
In fact, $\beta(d,\alpha)$ is a decreasing function in $d$. We note from the proof of the theorem that there is a tradeoff between $d$ and the probability of successful recovery. As such, the parameter $d$ has to be chosen judiciously. 
In addition, condition \eqref{eq:thmval} requires the stubborn agents to be sufficiently influential to the non-stubborn agents. \vspace{-.0cm}

\vspace{-.0cm}
\section{Randomized Models with Stubborn Agents} \label{sec:rand} \vspace{-.0cm}
% fluctuation is interesting to describe...
This section considers a general model of \eqref{eq:opn} with randomized opinion exchange where ${\bm W}(t)$ is time varying and i.i.d.~with mean $\overline{\bm W}$.
Under this setting, the limit equation \eqref{eq:limit_eq} only holds in expectation, i.e.,\vspace{-.0cm}
\begin{equation}
\begin{array}{l}
\displaystyle \lim_{t \rightarrow \infty} ( {\bm I} - \overline{\bm D}) \mathbb{E} \{ {\bm x}_{[n] \setminus [n_s]} (t;s) | {\bm x}(0;s)  \} \vspace{.05cm} \\
\displaystyle ~~~ = \lim_{t \rightarrow \infty} \overline{\bm B} \mathbb{E}\{ {\bm x}_{[n_s]} (t;s) | {\bm x}(0;s) \}. \vspace{-.0cm}
\end{array}
\end{equation}
Notice that the expectation is taken over the ensemble of \emph{sample paths} of $\{ {\bm W}(t) \}_{t}$. In practice, computing the expectation requires the social network to `repeat' the discussion on the same issue. Obtaining $\lim_{t \rightarrow \infty} \mathbb{E} \{ {\bm x} (t;s) | {\bm x}(0;s) \}$ can be difficult in terms of implementation.

Interestingly, it has been observed that in a randomized opinion exchange model, the introduction of stubborn agents leads to a behavior known as \emph{opinion fluctuation} \cite{Acemoglu2013,bianchi2012}.\vspace{.0cm}
\begin{observation} \label{obs:fluctuate} 
If $n_s \geq 2$ and the opinion exchange model is random, then ${\bm x}(t+1;s) \neq {\bm x}(t;s)$ almost surely.\vspace{.0cm}
\end{observation}

Our goal is to derive an estimator for $\lim_{t \rightarrow \infty} \mathbb{E} \{ {\bm x} (t;s) | {\bm x}(0;s) \}$ that relies on the samples from a single issue $s$ only. From Observation~\ref{obs:fluctuate}, a natural design is to consider an estimator that averages over the temporal samples, i.e., Eq.~\eqref{eq:sample}. We have \cite{Kay1993}:\vspace{.1cm}
\begin{theorem} \label{thm:asymp}
Consider the estimator in \eqref{eq:sample} with the sampling set ${\cal T}_s$. Denote $\overline{\bm x}(\infty; s) \triangleq \lim_{t \rightarrow \infty} \mathbb{E} \{ {\bm x}(t; s) | {\bm x}(0;s) \} = \overline{\bm W}^\infty {\bm x}(0;s)$ and assume that $\| \overline{\bm D} \|_2 = \mathbb{E} \{ \| {\bm D}(t) \|_2 \}$. If $T_o \rightarrow \infty$, then
\begin{enumerate}
\item the estimator \eqref{eq:sample} is unbiased: \vspace{-.1cm}
\begin{equation}
%\lim_{ |{\cal T}_s| \rightarrow \infty }
\mathbb{E} \{ \hat{\bm x}({\cal T}_s; s) | {\bm x}(0;s)\} =  \overline{\bm x}(\infty; s). \vspace{-.1cm}
\end{equation}
\item 
%If $T_o \rightarrow \infty$,
% and $\min_{t_i,t_j \in {\cal T}_s, i\neq j} | t_i - t_j | \rightarrow \infty$, 
 the estimator \eqref{eq:sample} is consistent: \vspace{-.1cm}
\begin{equation} \label{eq:est_cons}
\lim_{ |{\cal T}_s| \rightarrow \infty } \mathbb{E} \{ \| \hat{\bm x}({\cal T}_s; s) - \overline{\bm x}(\infty; s) \|_F^2 | {\bm x}(0;s) \} = 0.\vspace{.1cm}
\end{equation}
\end{enumerate}
\end{theorem}
%Notice that for the estimator to be consistent, we require a large number of samples to be taken. 
%Theorem~\ref{thm:asymp} indicates that the same estimator in \eqref{eq:sample} produces a reliable estimate of $\overline{\bm x}(\infty;s)$. 
It follows that the method in Section~\ref{sec:ssi} can be applied.

Note that the convergence of \eqref{eq:sample} is related to the ergodicity of the random process \eqref{eq:opn}.
For instance, \cite{Ravazzi2015} has studied the convergence of the ergodic mean of an opinion exchange model with external input. 
To our knowledge, our result is the first for randomized opinion exchange with stubborn agents.

\subsection{Proof of Theorem~\ref{thm:asymp}}
We first prove that the estimator is unbiased. Consider the following chain:
\begin{equation}
\begin{array}{l}
\displaystyle \mathbb{E} \{ \hat{\bm x}({\cal T}_s; s) | {\bm x}(0;s) \} = \frac{1}{|{\cal T}_s|} \sum_{t_i \in {\cal T}_s} \mathbb{E} \{ {\bm y}(t_i;s) | {\bm x}(0;s) \} \\
\displaystyle ~~=  \frac{1}{|{\cal T}_s|} \sum_{t_i \in {\cal T}_s} \overline{\bm W}^{t_i} {\bm x}(0;s) = \overline{\bm W}^\infty {\bm x}(0;s), \vspace{-.3cm}
\end{array}
\end{equation}
where we have used the fact that $T_o \rightarrow \infty$ and $t_i \geq T_o$ for all $t_i$ in the last equality.

Next, we prove that the estimator is asymptotically consistent, i.e., \eqref{eq:est_cons}. Without loss of generality, we let $t_1 < t_2 < \ldots < t_{|{\cal T}_s|}$ as the sampling instances. The following shorthand notation will be useful:
\begin{equation} \label{eq:phi_def}
\bm{\Phi}(s,t) \triangleq \bm{W}(t) \bm{W}(t-1) \ldots \bm{W}(s+1) \bm{W}(s),
\end{equation}
where $t \geq s$ and $\bm{\Phi}(s,t)$ is a random matrix.
Our proof involves the following lemma:
\begin{lemma} \label{lemma:phi}
When $|t-s| \rightarrow \infty$, the random matrix $\bm{\Phi}(s,t)$ converges almost surely to the following:
\begin{equation}
\lim_{ |t-s| \rightarrow \infty} \bm{\Phi}(s,t) = \left(
\begin{array}{cc}
{\bm I} & {\bm 0} \\
{\bm B}(s,t) & {\bm 0}
\end{array}
\right),
\end{equation}
where ${\bm B}(s,t) = \sum_{q=s}^t ( {\bm D}(t) \ldots {\bm D}(q) ) {\bm B}(q)$ is bounded almost surely. 
%Notice that lower diagonal of the random matrix is zero.
\end{lemma}
%\emph{Proof:}  See Appendix~\ref{app:as}. 
%\hfill \textbf{Q.E.D.}
The proof is in Appendix~\ref{app:as}. We consider the following:
\begin{equation}
\begin{array}{l}
\displaystyle \mathbb{E} \{ \| \hat{\bm x}({\cal T}_s; s) - \overline{\bm x}(\infty;s) \|_F^2 | {\bm x}(0;s) \} = \vspace{.1cm} \\
\displaystyle ~~= \mathbb{E} \Big\{ \Big\| \frac{1}{|{\cal T}_s|} \sum_{t_i \in {\cal T}_s} \big( {\bm y}(t_i;s) - \overline{\bm x}(\infty;s) \big) \Big\|_F^2 | {\bm x}(0;s) \Big\}. \vspace{-.3cm}
\end{array}
\end{equation}
Recall that ${\bm y}(t_i;s) = {\bm x}(t_i;s) + {\bm n}(t_i;s)$ and the noise term ${\bm n}(t_i;s)$ is independent of ${\bm W}(t)$ for all $t$. The above expression reduces into:
\begin{equation}
\begin{array}{l}
 \mathbb{E} \Big\{ \Big\| \frac{1}{|{\cal T}_s|} \sum_{t_i \in {\cal T}_s} \big( {\bm x}(t_i;s) - \overline{\bm x}(\infty;s) \big) \Big\|_F^2 | {\bm x}(0;s) \Big\} \\
 \hfill + \mathbb{E} \Big\{ \Big\| \frac{1}{|{\cal T}_s|} \sum_{t_i \in {\cal T}_s} {\bm n}(t_i;s) \Big\|_F^2 \Big\}.
\end{array}
\end{equation}
It is easy to check that the latter term vanishes when $|{\cal T}_s| \rightarrow \infty$. We thus focus on the former term.
\begin{equation}
\begin{array}{l}
\displaystyle \mathbb{E} \Big\{ \Big\| \frac{1}{|{\cal T}_s|} \sum_{t_i \in {\cal T}_s} \big( {\bm x}(t_i;s) - \overline{\bm x}(\infty;s) \big) \Big\|_F^2 | {\bm x}(0;s) \Big\} \\
\displaystyle = \frac{1}{|{\cal T}_s|^2} \mathbb{E} \Big\{ \Big\| \sum_{t_i \in {\cal T}_s} \big( \bm{\Phi}(0,t_i) - \overline{\bm W}^\infty \big) {\bm x}(0;s)  \Big\|_F^2 \Big\} \\
\displaystyle = \frac{1}{|{\cal T}_s|^2} \mathbb{E} \big\{ {\rm Tr} \big( \bm{\Xi} {\bm x}(0;s) {\bm x}(0;s)^T \big) \big\},
\end{array}
\end{equation}
where
\begin{equation}
\bm{\Xi} = \sum_{t_j \in {\cal T}_s} \big( \bm{\Phi}(0,t_j) - \overline{\bm W}^\infty \big)^T \sum_{t_i \in {\cal T}_s} \big( \bm{\Phi}(0,t_i) - \overline{\bm W}^\infty \big).
\end{equation}
Expanding the above product yields two groups of terms --- when $t_i = t_j$ and when $t_i \neq t_j$. 
When $t_i = t_j$, using $T_o \rightarrow \infty$ and  Lemma~\ref{lemma:phi}, it is straightforward to show that:
\begin{equation} \label{eq:mat_bd}
\begin{array}{l}
\displaystyle \| \mathbb{E} \big\{ \big( \bm{\Phi}(0,t_i) - \overline{\bm W}^\infty \big)^T \big( \bm{\Phi}(0,t_i) - \overline{\bm W}^\infty \big) \big\} \| \leq C,
\end{array}
\end{equation}
for some constant $C < \infty$. As a matter of fact, we observe that the above term will not vanish at all. This is due to Observation~\ref{obs:fluctuate}, the random matrix $\bm{\Phi}(0,t_i)$ does not converge in mean square sense.

For the latter case, we assume $t_j > t_i$. We have
\begin{equation}
\begin{array}{l}
\displaystyle \big( \bm{\Phi}(0,t_j) - \overline{\bm W}^\infty \big)^T \big( \bm{\Phi}(0,t_i) - \overline{\bm W}^\infty \big) \vspace{.1cm} \\
\displaystyle = \big( \bm{\Phi}(t_i+1,t_j) \bm{\Phi}(0,t_i) - \overline{\bm W}^\infty \big)^T \big( \bm{\Phi}(0,t_i) - \overline{\bm W}^\infty \big).
\end{array}
\end{equation}
Taking expectation of the above term gives:
\begin{equation} \label{eq:inter2}
\begin{array}{l}
\displaystyle \mathbb{E} \big\{ \big( \bm{\Phi}(0,t_i) - \overline{\bm W}^\infty \big)^T \overline{\bm W}^{t_j-t_i} \big( \bm{\Phi}(0,t_i) - \overline{\bm W}^\infty \big) \big\},
\end{array}
\end{equation}
where we have used the fact that $\bm{\Phi}(t_i+1,t_j)$ is independent of the other random variables in the expression and $\overline{\bm W}^{\infty} \overline{\bm W}^\ell = \overline{\bm W}^{\infty}$ for any $\ell \geq 0$.
Now, notice that 
\begin{equation}
\overline{\bm W}^{t_j-t_i} = \overline{\bm W}^{\infty} + {\cal O}(\lambda^{t_j-t_i}),
\end{equation}
for some $0 < \lambda \triangleq \lambda_{max}(\overline{\bm D}) < 1$. This is due to the fact that $\overline{\bm D}$ is sub-stochastic. 

As $T_o \rightarrow \infty$ and by invoking Lemma~\ref{lemma:phi}, the matrix $(\bm{\Phi}(0,t_i) - \overline{\bm W}^\infty)$ has almost surely \emph{only} non-empty entries in the lower left block. Through carrying out the block matrix multiplications and using the boundedless of $\bm{\Phi}(0,t_i)$, it can be verified that 
%the norm of the expectation in \eqref{eq:inter2} is upper bounded by ${\cal O}(\lambda^{t_j-t_i})$.
\begin{equation} \label{eq:inter2}
\begin{array}{l}
\displaystyle \big\| \mathbb{E} \big\{ \big( \bm{\Phi}(0,t_j) - \overline{\bm W}^\infty \big)^T  \big( \bm{\Phi}(0,t_i) - \overline{\bm W}^\infty \big) \big\} \big\| \displaystyle \leq {\cal O}(\lambda^{t_j-t_i}).
\end{array}
\end{equation}

%Consequently,  the above term vanishes as $|t_j - t_i| \rightarrow \infty$.

Combining these results, we can show
\begin{equation} \label{eq:final_bd}
\begin{array}{l} 
\displaystyle \frac{ \mathbb{E} \big\{ {\rm Tr} \big( \bm{\Xi} {\bm x}(0;s) {\bm x}(0;s)^T \big) \big\} }{|{\cal T}_s|^2} \leq \frac{{C'}}{|{\cal T}_s|} \Big( \sum_{i=0}^{|{\cal T}_s|-1} \lambda^{\min_{k} |t_{k+i} - t_k|}  \Big),
\end{array}
\end{equation}
for some $C' < \infty$. Notice that $\min_{k} |t_{k+i} - t_k| \geq i$ and the terms inside the bracket can be upper bounded by the geometric series $\sum_{i=0}^{|{\cal T}_s|-1} \lambda^i < \infty$.  
%\begin{equation}
%\frac{1}{|{\cal T}_s|^2} \mathbb{E} \big\{ {\rm Tr} \big( \bm{\Xi} {\bm x}(0;s) {\bm x}(0;s)^T \big) \big\} \leq \frac{C}{|{\cal T}_s|},
%\end{equation}
Consequently, the mean square error goes to zero as $|{\cal T}_s| \rightarrow \infty$. The estimator \eqref{eq:sample} converges in the mean square sense and is thus consistent.

\begin{remark}
From \eqref{eq:final_bd}, we observe that the upper bound on mean square error can be minimized by maximizing $\min_{i,j, i \neq j} |t_i - t_j|$. 
When the samples ${\bm y}({\cal T}_s;s)$ are taken from a finite  interval $[T_{max}] \setminus [T_o]$, $T_{max} < \infty$ and $|{\cal T}_s| < \infty$, the best estimate can be obtained by using sampling instances that are drawn uniformly from $[T_{max}] \setminus [T_o]$. 

%As we have $|{\cal T}_s| < \infty$ in practice, the best strategy is to collect the samples is to ensure that any $t_i, t_j$ are taken at far away distance in time.  
\end{remark}

\section{Numerical Results}
This section provides numerical results for the performance of SSI. Two simple scenarios on the static and randomized model are considered.

The social graph $G = (V,E)$ is generated as an Erdos-Renyi graph with connectivity  $p_e = 0.15$. We fix the number of normal agents at $n-n_s=50$. We assume $m = 1$ and the number of issues to be discussed is $K = 100$. 
%The condition \eqref{eq:nec_cond}  implies that $n_s \geq 13$ is needed in this case.
For the static model, the trust matrix $\overline{\bm{W}}$ is first generated with uniformly distributed entries, which are then normalized to satisfy row-stochasticity as well as the sparsity pattern according to $E$; cf.~Assumption~\ref{assume:E}.
For the randomized model, we have adopted the randomized broadcast gossip exchange model in \cite{Aysal2009}. In particular, at each time, a random agent wakes up and broadcast his/her opinion to the neighbors. The neighbors then mix the opinion with the weight $\gamma = 1/2$; see \cite{Aysal2009}.
%Lastly, we assume that $D_{ii} \geq D_{ij}$ for all $j$, i.e., an agent  has always a higher self trust than his/her trust on the neighbors.

%In solving \eqref{eq:ssi}, we notice that there is an ambiguity

%We notice that there is an ambiguity in determining the self trust

\begin{figure}[t]
\centerline{\resizebox{.23\textwidth}{!}{
\includegraphics{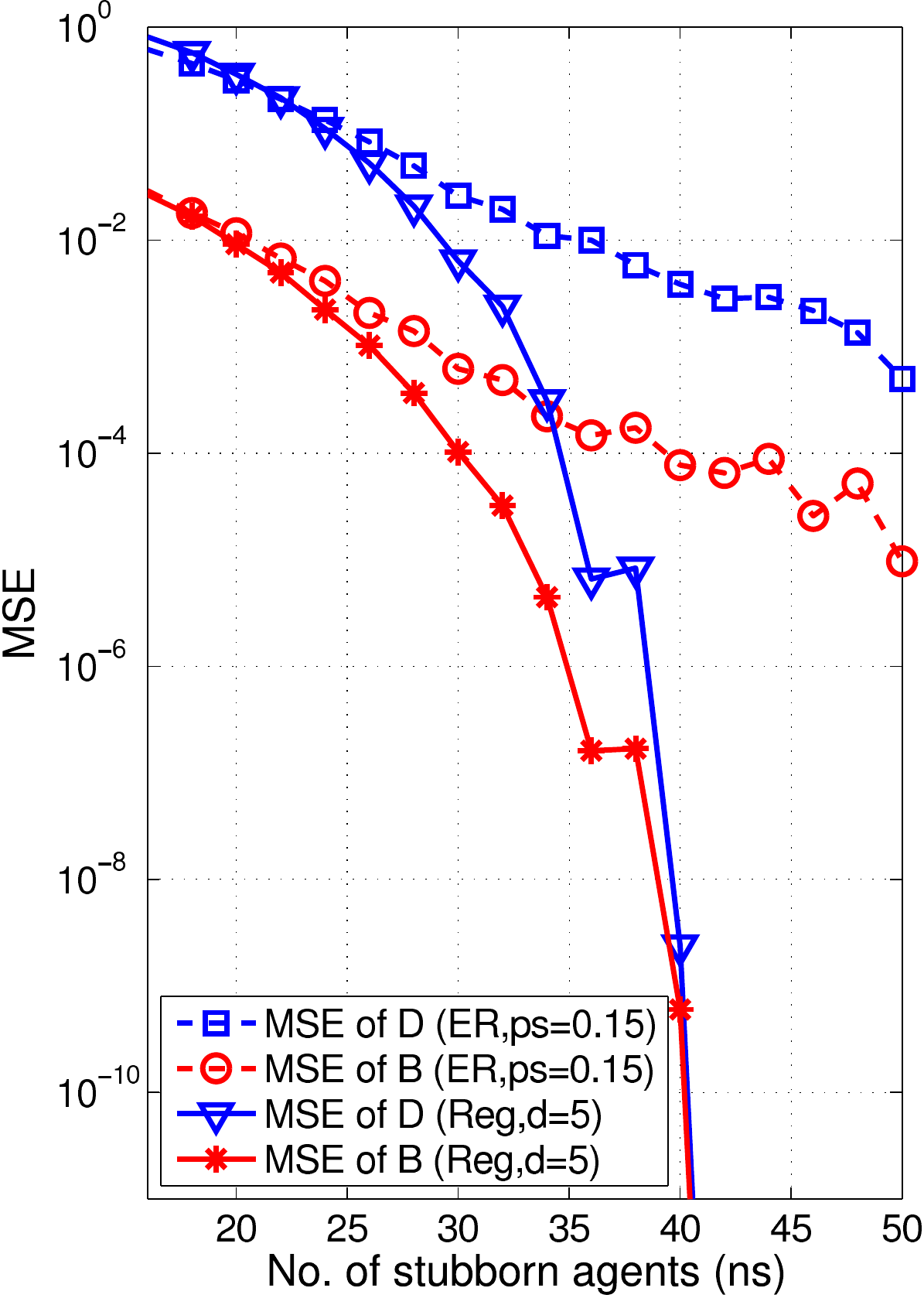}} \resizebox{.225\textwidth}{!}{
\includegraphics{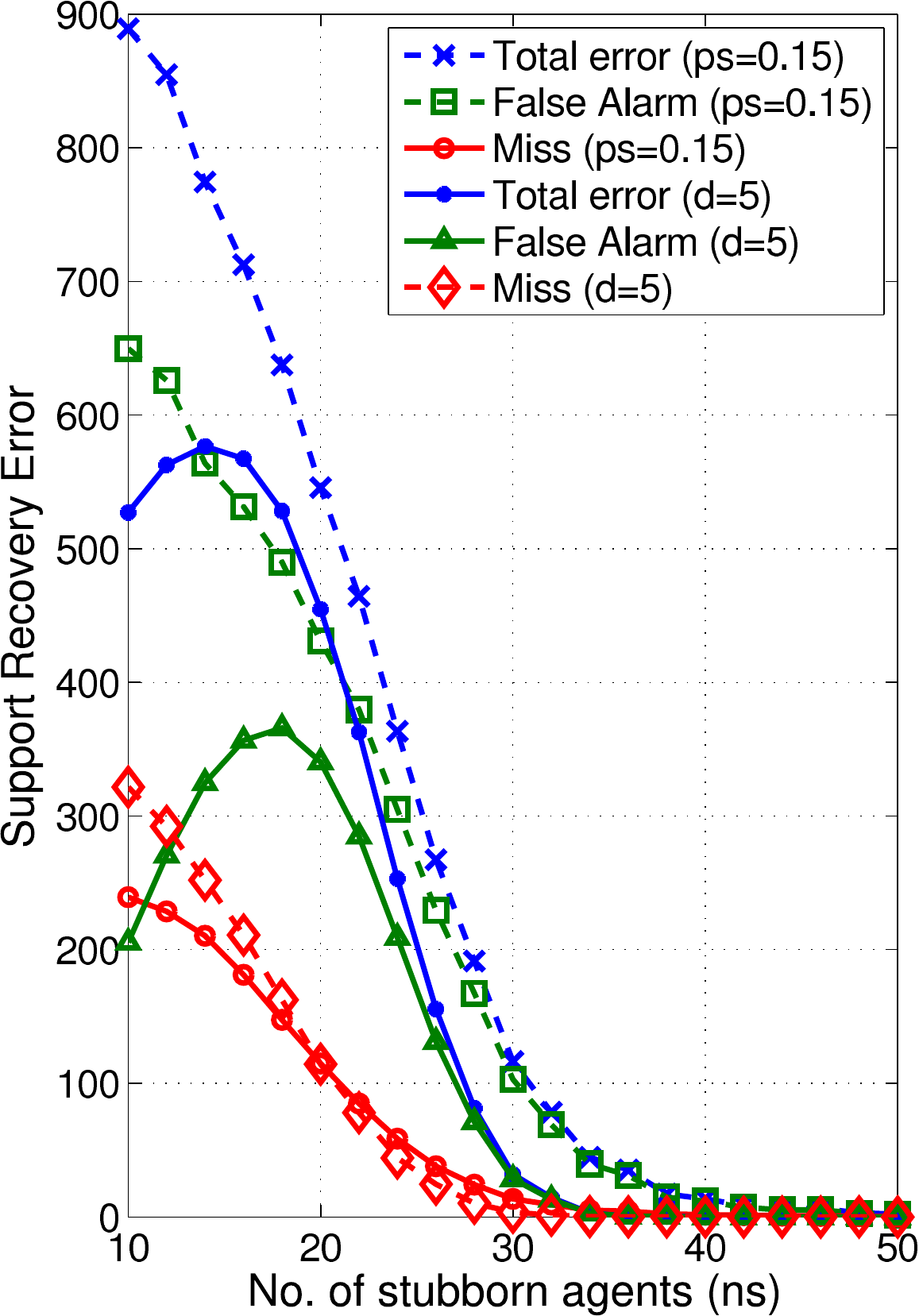}}} \vspace{-.2cm}
\caption{Average performance of social system identification under the static model: (Left) the relative MSE in estimating $\overline{\bm D}^r$ and $\overline{\bm B}^r$~(Right)~the support recovery error --- $| \{ ij : \overline{D}_{ij} \neq 0, \hat{D}_{ij} = 0, {\rm or}, \overline{D}_{ij} = 0, \hat{D}_{ij} \neq 0\} |$.
} \label{fig:mse} \vspace{-.4cm}
\end{figure}

In light of Lemma~\ref{lem:amb}, 
we compare the relative trust that an agent has on his/her neighbors. In particular, we evaluate the error in estimating the relative trust matrix ${\overline{\bm D}}^r, \overline{\bm B}^r$ as $\overline{D}_{ij}^r = \overline{D}_{ij} / (1 - \overline{D}_{ii})$, $\overline{B}_{ij}^r = \overline{B}_{ij} / (1 - \overline{D}_{ii})$ and $\overline{D}_{ii}' = 0$ for all $i$. The normalized mean square error (MSE) for $\overline{\bm D}$ is $\sum_{i,j} (\hat{D}_{ij} - \overline{D}_{ij}^r)^2 / (\sum_{i,j}(\overline{D}_{ij}^r)^2)$ (and similarly for $\overline{\bm B}$).

We first evaluate the SSI performance under the static model with Monte-Carlo simulation. For each $n_s$, we average over $100$ instances of social graphs to evaluate the normalized MSE. The noise $\sigma^2$ is assumed to be zero. As such, we can set $T_o = 10^4$ and $|{\cal T}_s| = 1$ for the estimator \eqref{eq:sample}. We compare the normalized MSE in terms of the relative trust matrices, against the number of stubborn agents $n_s$. The simulation results are shown in Fig.~\ref{fig:mse}. As seen, the system identification performance gradually improves as $n_s$ grows. 
Moreover, the performance is significantly better when the subgraph between stubborn and non-stubborn agents is constructed as a random regular bipartite graph, cf.~Theorem~\ref{thm:cs}. 
Notice that the theorem's condition requires $n_s \geq 38$. 
%In addition, we  observe a sharp improvement in support recovery error when $n_s \geq 20$.

\begin{figure}[t]
\centerline{\resizebox{.225\textwidth}{!}{
\includegraphics{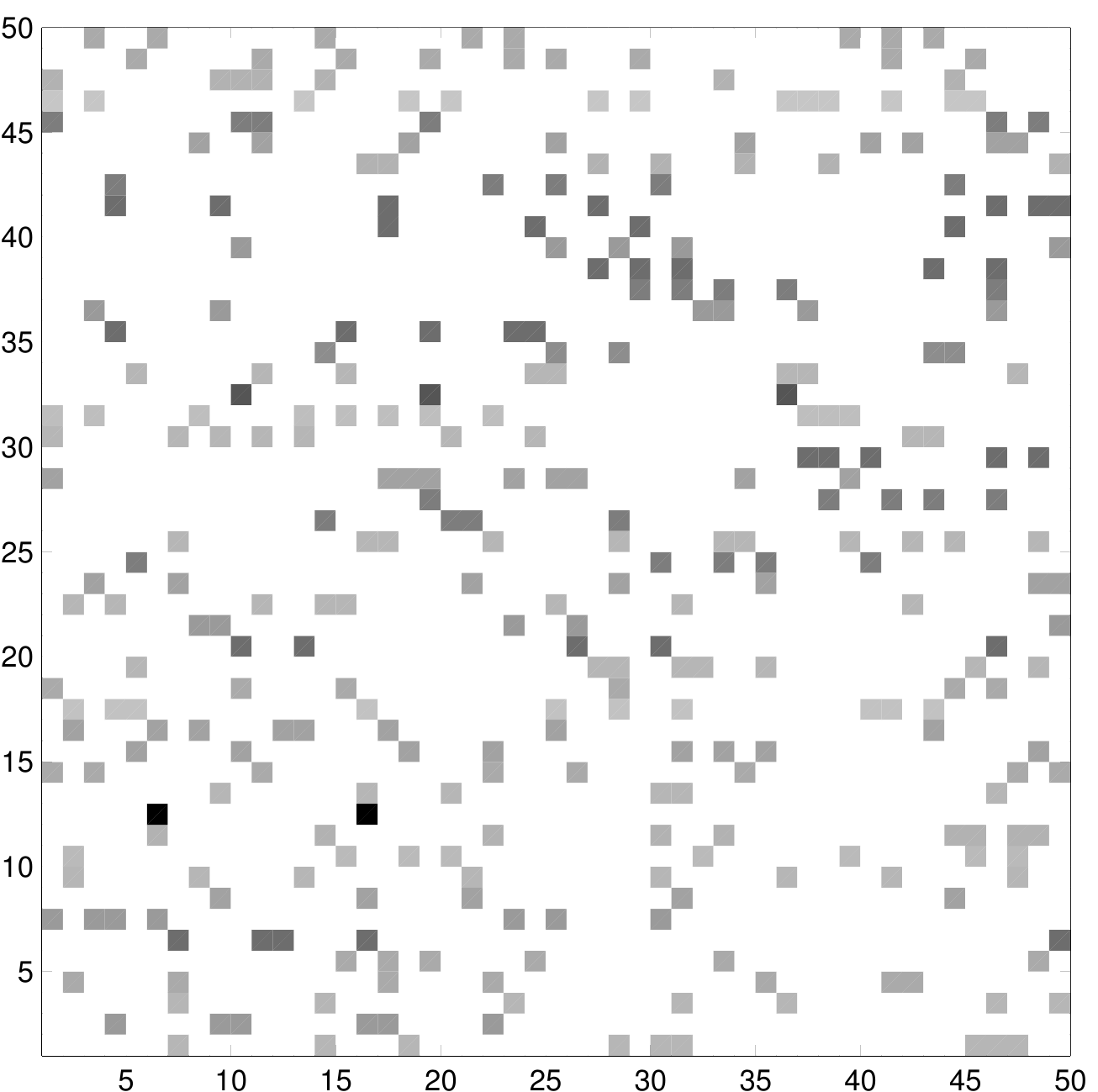}}~\resizebox{.225\textwidth}{!}{
\includegraphics{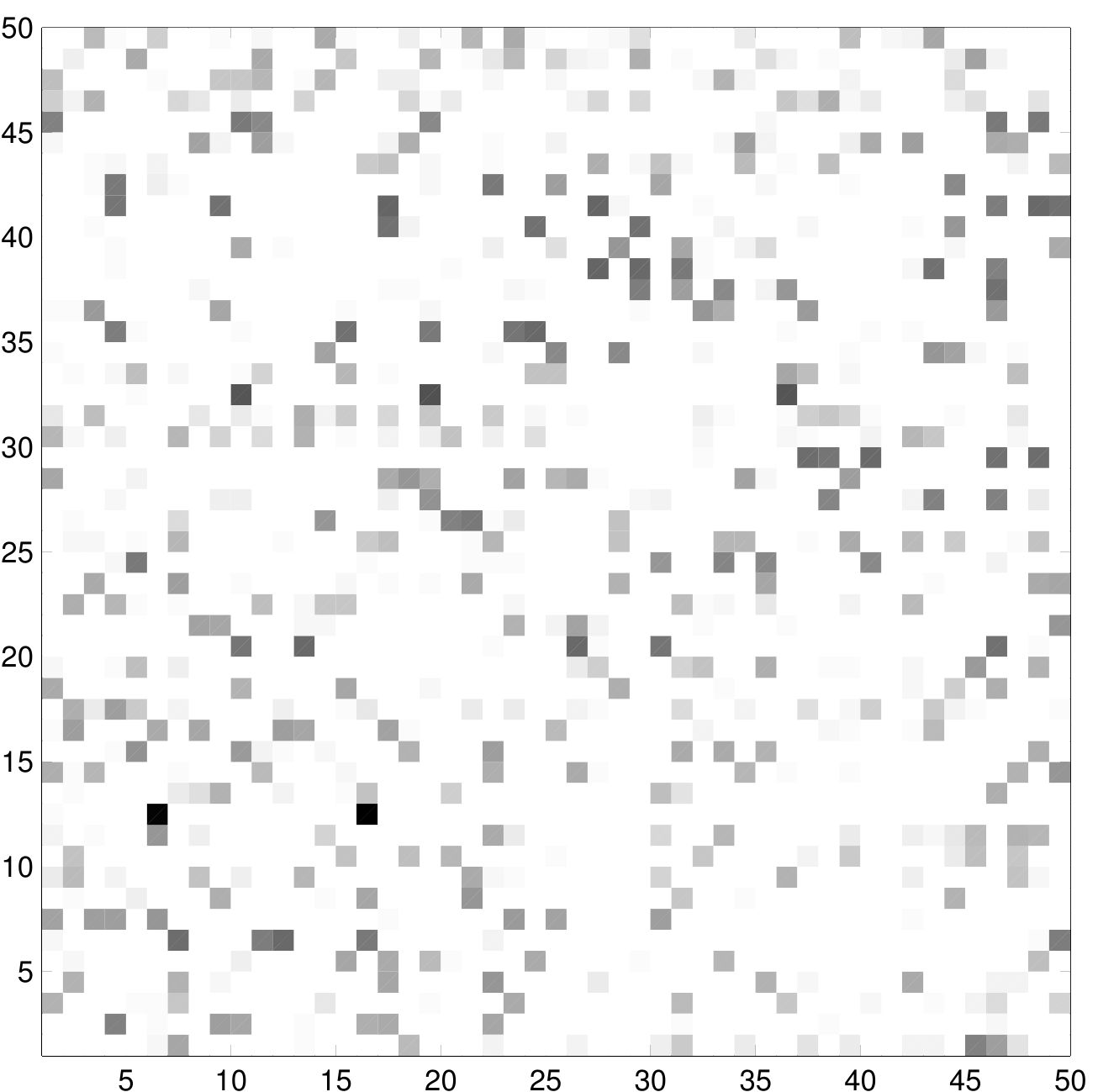}}} \vspace{-.1cm}
\caption{Identifying a social system under the randomized gossip exchange model: (Left) the normalized trust matrix $\overline{\bm D}^r$ of the actual social system.~(Right)~the normalized estimated trust matrix $\hat{\bm D}^r$. The darkness of the dot represents the amount of trust between a pair of agents.
} \label{fig:social} \vspace{-.5cm}
\end{figure}

The next simulation example considers the case with random opinion exchanges. We focus on one instance of the social graph generated and we set the number of stubborn agents to $n_s = 30$. We conduct the test when $\Omega_{\overline{\bm B}}$ is generated as a random non-regular bipartite graph of connectivity $p_s = 0.15$. For the estimator proposed in Section~\ref{sec:rand}, we set $T_o = 10^3$ and $|{\cal T}_s| = 3 \times 10^3$, where sampling instances $t_i$ in ${\cal T}_s$ is uniformly drawn from $\{10^3+1,\ldots, 10^5\}$. The  noise variance is $\sigma^2 = 10^{-4}$ and we set $\epsilon = \sqrt{K} \times 1.65 \times 10^{-2}$ in \eqref{eq:ssi}. The estimated social system is depicted in Fig.~\ref{fig:social}. Notice that in this case, the normalized MSE is evaluated as $7.01 \times 10^{-2}$.

We observe that the identified social system is close to the actual social system. However, some links with weak trusts can also be found in the estimate $\hat{\bm D}^r$. This is possibly an artefact from the estimation of $\overline{\bm x}(\infty;s)$ using \eqref{eq:sample}.

\section{Conclusions}
We have defined the SSI problem for identifying both the social graph and mutual trusts between individuals in social networks. The system identification is achieved via the inclusions of stubborn agents and conditions for identifiability are proven. We have  proposed a consistent estimator for the ensemble mean opinions in randomized gossip model.
%Our preliminary simulation results show that a social system can be identified with a reasonable performance.

This work paves a key stone towards developing a \emph{Social Radar} that estimates the relative influence between individuals. Future directions will include developing an efficient and parallelizable solution method for solving \eqref{eq:ssi_r} and tightening the necessary condition for identifiability.

\appendix
\subsection{Proof of Lemma~\ref{lemma:phi}} \label{app:as}
We first establish the almost sure convergence of ${\bm D}(t) {\bm D}(t-1) \ldots {\bm D}(s)$ to ${\bf 0}$. Define
\begin{equation}
\beta(s,t) \triangleq \| {\bm D}(t) {\bm D}(t-1) \ldots {\bm D}(s) \|_2,
\end{equation}
and observe the following chain
\begin{equation}
\begin{array}{l}
\mathbb{E} \{ \beta(s,t) | \beta(s,t-1),...,\beta(s,s) \} \vspace{.05cm} \\
% = \mathbb{E} \{ \| {\bm D}(t) {\bm D}(t-1) \ldots {\bm D}(s) \|_2 | \beta(s,t-1) \}  \vspace{.05cm} \\
  \leq \mathbb{E} \{ \| {\bm D}(t) \|_2 \| {\bm D}(t-1) \ldots {\bm D}(s) \|_2 | \beta(s,t-1) \}  \vspace{.05cm} \\
  = \mathbb{E} \{ \| {\bm D}(t) \|_2 \} \beta(s,t-1) \leq c \beta(s,t-1),
\end{array}
\end{equation}
where $c = \| \overline{\bm D} \|_2 < 1$ due to Assumption~\ref{assume:b}. The almost sure convergence of $\beta(s,t)$ follows from \cite[Lemma~7]{polyak87}. Now, expanding the multiplication \eqref{eq:phi_def} yields:
 \begin{equation}
\bm{\Phi}(s,t) = \left(
\begin{array}{cc}
{\bm I} & {\bm 0} \\
{\bm B}(s,t) & {\bm D}(t) \ldots {\bm D}(s)
\end{array}
\right).
\end{equation}
The desired result is achieved by observing ${\bm D}(t) \ldots {\bm D}(s) \rightarrow {\bf 0}$ and ${\bm B}(s,t)$ is bounded almost surely.

\bibliographystyle{IEEEtran}
%\small
%\bibliographystyle{unsrt}
\bibliography{social}
%\bibliography{bibliography}
\end{document}